# On the influence of annealing on the compositional and crystallographic properties of sputtered Li-Al-O thin films

Florian Lourens[a], Detlef Rogalla[b], Ellen Suhr[a], Alfred Ludwig[a,] *

[a] *Materials Discovery and Interfaces, Institute for Materials, Ruhr University Bochum, Universitätsstrasse 150, 44780 Bochum, Germany*

[b] *RUBION, Central Unit for Ion Beams and Radionuclides, Ruhr University Bochum*

*\* Corresponding author: alfred.ludwig@rub.de*

Keywords: Magnetron sputtering, combinatorial materials science, Li-aluminate

## Abstract

A Li-Al-O thin film materials library, deposited by inert magnetron sputtering and post-deposition annealing in $O_2$ atmosphere, was used to study the effects of different annealing temperatures (300 – 850°C) and durations (1 min to 7 h) on crystallinity and composition of the films. XPS depth profiling revealed inhomogeneous compositional depth profiles with Li contents increased toward the film surface and Al contents toward the film-substrate interface. These depth profiles were confirmed by a combination of RBS and D-NRA. At annealing temperatures of 550°C and higher, Li reacted with the Si substrate. At the same time, temperatures of 550°C and higher enabled the formation of crystalline $LiAlO_2$, whereas at lower temperatures, no crystalline Li-Al-O phases were detected with XRD. In contrast to conventional annealing in a tube furnace (3 – 7 h durations), rapid thermal annealing with fast heating/cooling rates of 10°C/min and durations of 1 – 10 min resulted in homogeneous depth profiles, while also leading to crystalline $LiAlO_2$.

**Introduction**

The system Li-Al-O comprises the ternary compounds $LiAl_5O_8$, $LiAlO_2$ and $Li_5AlO_4$ [1–4], which are of interest for various applications. Ordered $LiAl_5O_8$ doped with rare-earth or transition metal cations shows red luminescence properties [2, 5, 6]. α- and γ-$LiAlO_2$ also show luminescence properties [2]. γ-$LiAlO_2$ was used as substrate material for epitaxial GaN semiconductor films [7]. Coating $LiCoO_2$ cathodes with $LiAlO_2$ can improve Li-ion battery performance [8]. $Li_5AlO_4$ is interesting for $CO_2$ absorption [9, 10], as solid state electrolyte, as insulator layer in thin film transistors [11], and as tritium breeder material [12].

The ternary Li-Al-O phases show a remarkable variety of polymorphs, and the different structures are a matter of debate in literature. $LiAl_5O_8$ exhibits low- and high-temperature polymorphs with ordered inverse spinel ($P4_332$ or $P4_132$) and disordered spinel structures ($Fd3m$), with a transition temperature of 1298°C [4, 13, 14]. In [15], the $LiAl_5O_8$ polymorphs were reported differently to have spinel and primitive cubic structure, but without reference or explanation. $LiAlO_2$ is reported to form two or three polymorphs. High-temperature γ-$LiAlO_2$ is consistently reported to have tetragonal structure ($P4_12_12$) [2, 4, 14, 16]. In [2, 16–18], two polymorphs additionally to tetragonal γ are mentioned, namely hexagonal α (stable at < 400°C) and monoclinic β (stable at 400 to 800°C). Hexagonal α was also mentioned in [14], but without mentioning a third intermediate polymorph. In [4], low-temperature α-$LiAlO_2$ has a cubic structure ($I4_132$) and the α ↔ γ transition temperature is 750°C, also without mentioning an intermediate form. The two polymorphs of $Li_5AlO_4$ are low-temperature α (orthorhombic, $Pbca$), and high-temperature β (orthorhombic, $Pmmn$), with α ↔ β transition temperature of about 827°C [10–12].

Li-Al-O compounds in bulk and powder form were fabricated by various methods such as wet-chemical processes [19], solid-state reaction [10, 16], and sol-gel methods [2, 11, 20].

Crystalline single-phase films of α-LiAl$_5$O$_8$, γ-LiAlO$_2$ and β-Li$_5$AlO$_4$ with thicknesses of approximately 2 – 4 µm were prepared by laser chemical vapor deposition at temperatures of 1196 – 1553°C [14]. Caudron et al. reported the sputter deposition of Li-Al-O thin films (0.2 – 2 µm thickness) on alumina substrates using LiAl$_5$O$_8$ and LiAlO$_2$ sputter targets in Ar/O$_2$ gas mixture and no intentional substrate heating, which resulted in amorphous films [21]. The publication from Jafarpour et al. discusses the fabrication of Li-Al-O thin films (150 nm thickness) on glass and quartz substrates by Li (RF) and Al (DC) co-sputtering in reactive Ar/O$_2$ atmosphere [22]. They report the as-deposited films to be amorphous, whereas annealing at 950°C for 2 h caused the crystallization of γ-LiAlO$_2$.

In this manuscript we investigated the influence of annealing on elemental distribution in sputtered Li-Al-O thin films along the growth direction, which in many cases turns out to be inhomogeneous. However, this unfavorable phenomenon can be mitigated by rapid thermal annealing, which at the same time allows crystallization of the γ-LiAlO$_2$ phase.

**Methods**

A Li-Al-O thin film materials library (ML) was fabricated by magnetron co-sputtering in Ar atmosphere and post-deposition annealing in O$_2$ atmosphere. The used sputter system (Creavac, Germany) has a sputter-up configuration with two confocal cathodes 120° apart. To establish Li and Al composition gradients in the thin film ML, sputter targets of elemental Li (50.8 mm diameter, 99.7 % purity, Mateck) and elemental Al (101.6 mm diameter, 99.995 % purity, KJ Lesker) were co-sputtered for 2 h at 30 W DC (Li) and 80 W DC (Al). The relative target locations are schematically depicted in Fig. 1 a. A Si wafer (100 mm diameter) with native oxide was used as substrate. The base pressure of the sputter chamber prior to the deposition was 2.1 x 10$^{-4}$ Pa, and the deposition pressure was 1 Pa with an Ar (6.0, Praxair) flow of 50 SCCM. The first post-deposition annealing was carried out in the sputter chamber without exposing the as-deposited Li-Al film to ambient air. The annealing temperature was 300°C, the duration 3 h, and the pressure 1 Pa with an $O_2$ flow of 10 SCCM. In this initial state of the ML (after the first post-deposition annealing), 342 and 88 measurement areas (MAs, each 4.5 x 4.5 mm²) distributed evenly across the thin film were investigated with high-throughput X-ray diffraction (XRD) and X-ray photoelectron spectroscopy (XPS), respectively. One MA at the center of the ML was further investigated with XPS depth profiling. The ML was then broken into fragments, seven of which were used for different annealing treatments in a custom-built tube furnace and in a rapid thermal annealing system (RTA, Annealsys AS-One). A photograph of the fragmented ML is shown in Fig. 1 a, and the annealing conditions are summarized in Fig 1 b. Sample fragment VII (in its initial state) was additionally investigated with transmission electron microscopy (TEM). After the different annealing treatments, XPS depth profiling and XRD measurements were carried out. To ensure that the measured XPS depth profiles were not affected by preferential sputtering, six of the sample fragments were additionally investigated by a combination of Deuteron-induced nuclear reaction analysis (D-NRA) and Rutherford Backscattering Spectrometry (RBS).

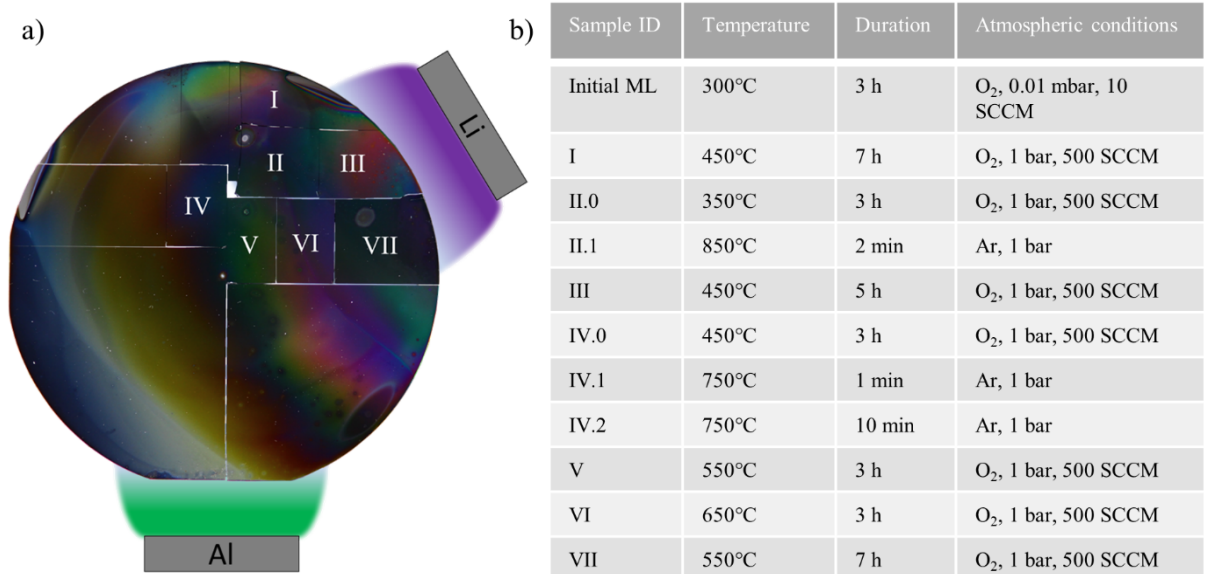

Figure 1: a) Photograph of the fragmented Li-Al–O thin film materials library (100 mm diameter). Roman numerals identify the differently annealed sample fragments. The locations of the Li and Al sputter targets are displayed (not to scale, arbitrary colors). b) Table 1 presents the parameters of the applied annealing treatments. The initial annealing was performed on the whole as-deposited materials library. Samples II and IV underwent multiple annealing treatments performed in numerical order according to the decimals of the sample IDs.

XPS measurements were carried out on a Kratos Axis Nova, using a monochromatic Al Kα X-ray source operating at 300 W (20 mA, 15 kV). A delay-line detector was used with pass energies of 160 eV and 20 eV for survey and high-resolution (Li 1s, Al 2p, O 1s, C 1s) scans, respectively. The electron emission angle was 0°, and the charge neutralizer was activated. An Ar$^+$ ion source with ion energies 4 keV, and a raster size of 1.5 x 1.5 mm$^2$ was used to perform the depth profiling. The analysis area was 300 x 700 μm$^2$ for surface measurements and 110 x 110 μm$^2$ for depth profiling. The measured spectra were charge-corrected with adventitious C-C at 284.8 eV. Quantification was done with the Kratos ESCApe software and its pre-defined relative sensitivity factors.

D-NRA and RBS were used in combination to measure light (Li, C, O) and heavy elements (Al, Si) with depth information. Both D-NRA and RBS were performed at the 4 MeV tandem accelerator facility RUBION (Ruhr-University Bochum). For RBS, a singly charged He beam with an energy of 3 MeV and a beam current of about 40 nA was used. A silicon surface barrier detector was placed at an angle of 160° with respect to the beam axis. The solid angle of the detector was 1.9 msrad. For the D-NRA measurements, a 1 MeV deuteron beam was directed onto the samples. Emerging protons from nuclear reactions with light elements were detected at an angle of 135° with respect to the beam axis. The detector covered a solid angle of 19 msrad and was shielded by a 6 μm Ni-foil to eliminate elastically scattered deuterons. The RBS and D-NRA spectra were analyzed using SIMNRA software [23].

A TEM cross-sectional sample (lamella) of sample fragment VII was prepared using a Helios NanoLab G4 CX FIB system, which is a dual beam system with an electron and a Ga-ion beam.

It is equipped with a micromanipulator (EasyLift, Oxford Instruments) and gas injection systems that were used for the deposition of protective Pt layers. The prepared lamella for TEM had an initial width of approximately 5 µm and a final thickness of less than 100 nm. The TEM images were recorded on a Jeol Neoarm aberration-corrected TEM operating at 200 kV.

XRD measurements were done on a Bruker D8 Discover with Bragg-Brentano geometry, using Cu Kα radiation collimated to 1 mm with a divergence of less than 0.007° and a VANTEC-500 area detector with a sample-to-detector distance of 149 mm. Each measurement comprised three stepwise recorded frames at 2θ of 25°, 45° and 65°, each covering 2θ increments of ±15°. XRD patterns were obtained by integrating the detector frames and subtracting the background signal. To identify the crystal structures of the present phases, the measured patterns were compared with reference patterns form the Inorganic Crystal Structure Database (ICSD) [24].

**Results and discussion**

The XPS mapping of the ML in its initial state revealed that the near-surface region of the film does not contain Al. Instead, it contains almost constant concentrations of Li (26 – 30 at. %), O (45 – 48 at. %) and C (25 – 27 at. %) without continuous composition gradients. The uniformity of the elemental distribution across the ML is highlighted in Fig. 2 a using a global color scale. Fig. 2 b shows the individual elemental composition gradients in more detail using individual color scales. The reason for this observation, which contradicts the expected result, is that within the XPS probing depth of around 10 nm, a reaction layer of $Li_2CO_3$ exists, paired with the usual surface contamination of adventitious C. This was identified by XPS peak fitting, as shown exemplary by the fits of one MA in Fig. 3. The C 1s spectra were fitted with components for adventitious C (C-C, C-H at binding energy BE = 284.8 eV, C-OH, C-O-C at BE = 286.3 eV, C=O at BE = 287.8 eV, O-C=O at BE = 288.7 eV, all with equal FWHM = 1.2 – 1.4 eV and G*L mix = 0.3) and carbonate species (BE = 289.6 – 289.7 eV, FWHM = 1.0 – 1.1 eV, G*L mix = 0.3). The Li 1s spectra were fitted with one component (BE = 55.0 – 55.2 eV, FWHM = 1.2 – 1.3 eV, G*L mix = 0.3). The O 1s spectra were fit with components for carbonate species (BE = 531.4 – 531.5 eV, FWHM = 1.3 – 1.4 eV, G*L mix = 0.2) and organic species as well as possible water (BE = 532.7 – 532.8 eV, FWHM = 1.4 – 1.5 eV, G*L mix = 0.2). The distances between the Li 1s peak and the Li$_2$CO$_3$ components of the C 1s and O 1s spectra are in agreement with [25], indicating a clear identification of Li$_2$CO$_3$.

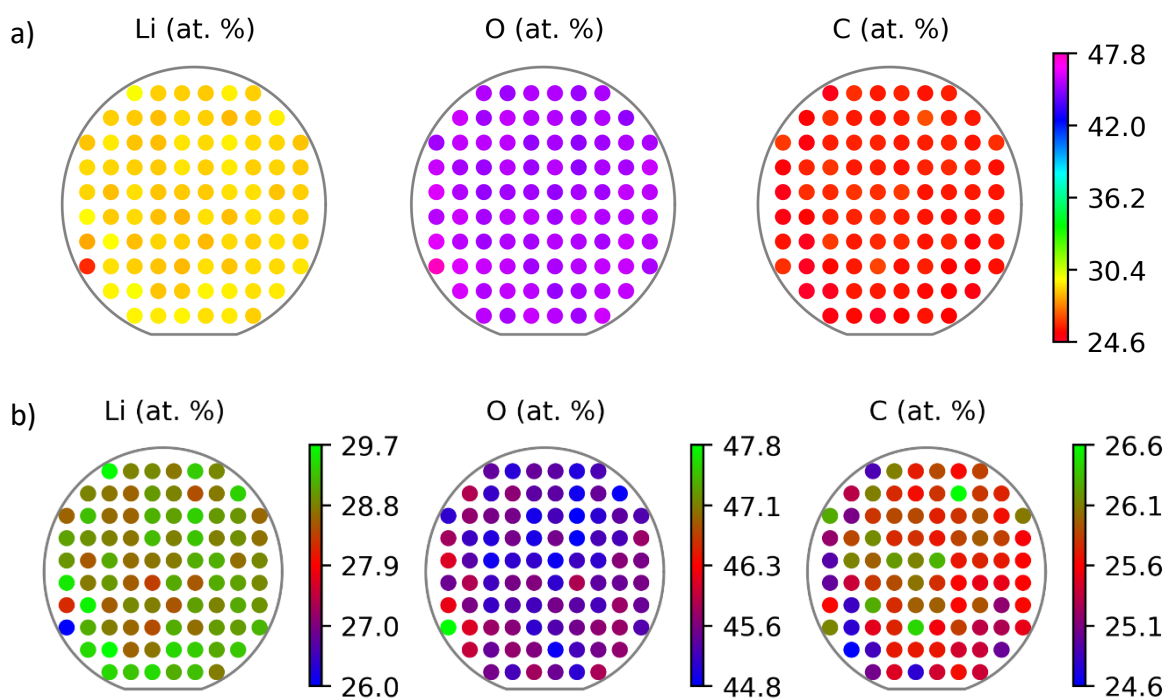

Figure 2: Color-coded elemental distribution of Li, O and C across the materials library in its initial state. The gray circle represents the contour of the substrate. a) Overview maps with a global color scale, highlighting the uniformity of the elemental distribution. b) Detailed elemental distribution maps with individual color scales.

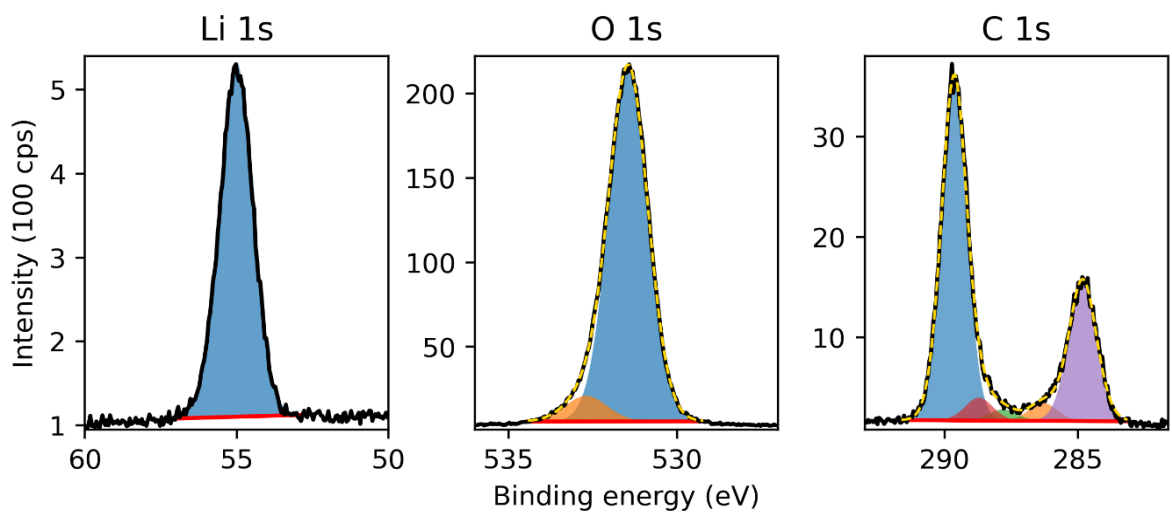

Figure 3: Exemplary XPS peak fitting results of one MA of the ML in its initial state of Li 1s, O 1s and C 1s. The black lines represent the measured spectra, the red lines the backgrounds, and the dashed yellow lines the fits. The colored areas are the fitted components.

The XPS depth profile of the central MA of the ML in its initial state reveals an inhomogeneous profile, as shown in Fig. 4 a. By removing adventitious C with the first etching step, the Li content increases from 32.9 to 45.6 at. %. Starting from 150 s etching time, the concentration of Li decreases while that of Al increases. As etching time increases, the C content steadily decreases while the O content remains relatively stable within the range of 46.3 – 56.7 at. %. The presence of C through the whole depth profile is undesired and considered a contamination. It can be ascribed to Li$_2$CO$_3$ based on the measured binding energy of C 1s of 290.2 – 290.3 eV. Starting from 540 s etching time, increasing Si content originating from the wafer substrate was detected. Annealing at 350 – 650°C for 3 – 7 h resulted in depth profiles (Fig. 4 b, d, e, g – h) that are qualitatively similar to the initial depth profile (Fig. 4 a). At 650°C (Fig. 4 j), the depth profile indicates a reaction between Li and the Si substrate, as the Li content increases with the Si content. This effect can also be observed for the two samples annealed at 550°C (Fig. 4 h, i), albeit more subtly. A reaction of Li and Si during annealing was also observed by Jafarpour et al. [22].

The depth profiles of the two RTA-annealed samples shown in Fig. 4 c and f are noticeably more homogeneous compared to the others. Both 850°C for 2 min (Fig. 4 c) and 750°C for 10 min (Fig. 4 f) resulted in evened-out Li and Al depth profiles compared to the previous sample states (Fig. 4 b and Fig. 4 e, respectively). After annealing at 850°C for 2 min, there is still some Li surface segregation (etching times 1 – 60 s). Also, Li reacted with the Si wafer (etching time > 360 s) and Si migrated into the film. These effects did not occur at 750°C for 10 min. Instead, the depth profile in Fig. 4 f is the most homogeneous one. At etching times of 0 – 300 s, i.e. up to when the Si signal was detected, the Li content is relatively stable, decreasing slightly from 12.4 to 8.4 at. %.

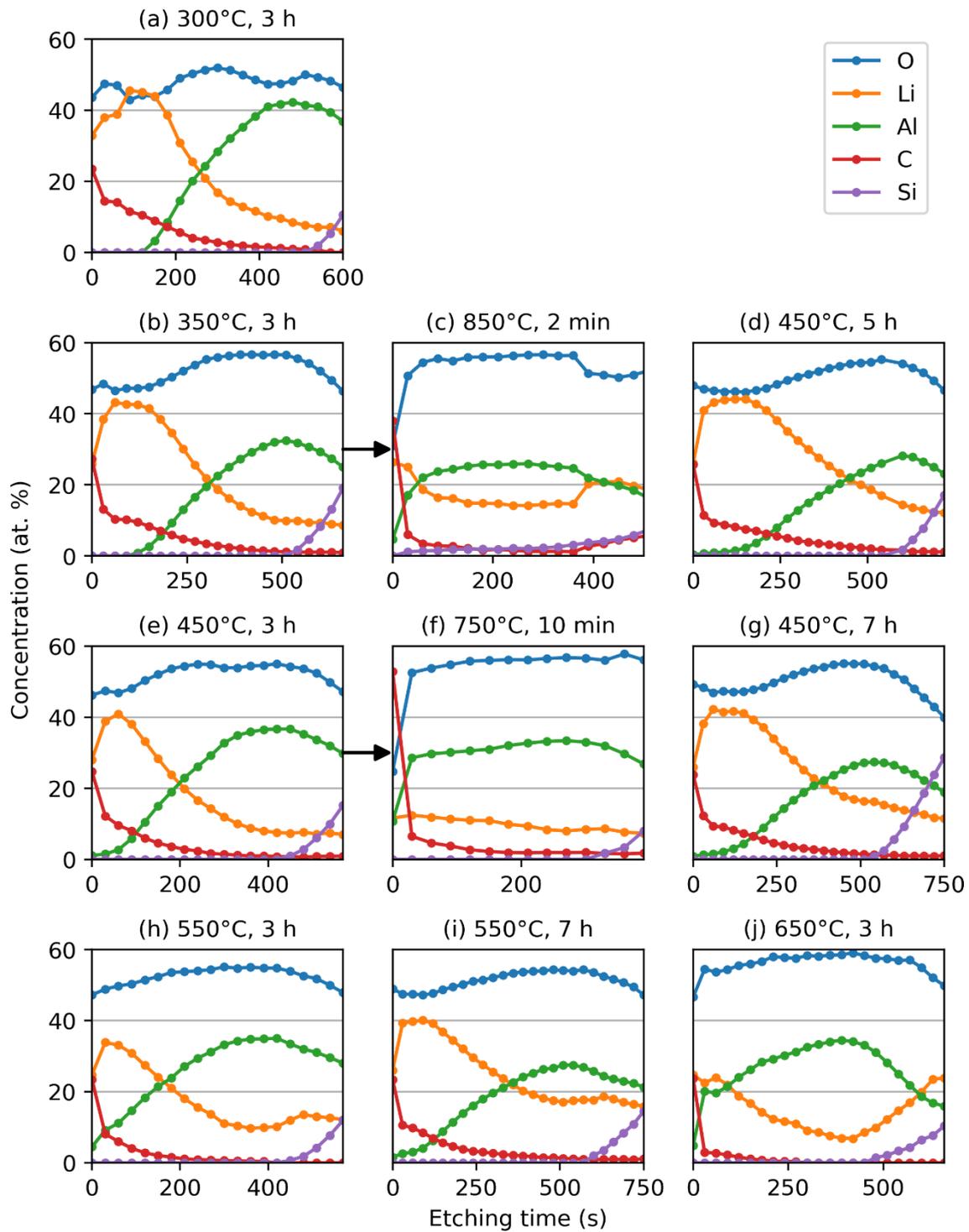

Figure 4: XPS depth profiles of the Li-Al-O thin films after different annealing treatments. The arrows point from the previous to the latest states of the same samples.

The Al content also remains relatively stable, increasing from 28.6 to 33.4 at. %, when the Al-depleted surface (at etching time 0 s) and the mixed film/wafer signal (at etching times > 300 s) are disregarded.

The D-NRA/RBS results of the six investigated samples shown in Fig. 5 confirm the XPS results of decreasing Li and increasing Al contents with increasing film depth. The initially increasing Li contents and C-rich surfaces observed with XPS were not detected by D-NRA/RBS, probably due to the limited depth resolution of the latter. A qualitative difference is that D-NRA/RBS indicates elevated C contents at the film-substrate interface, which XPS does not.

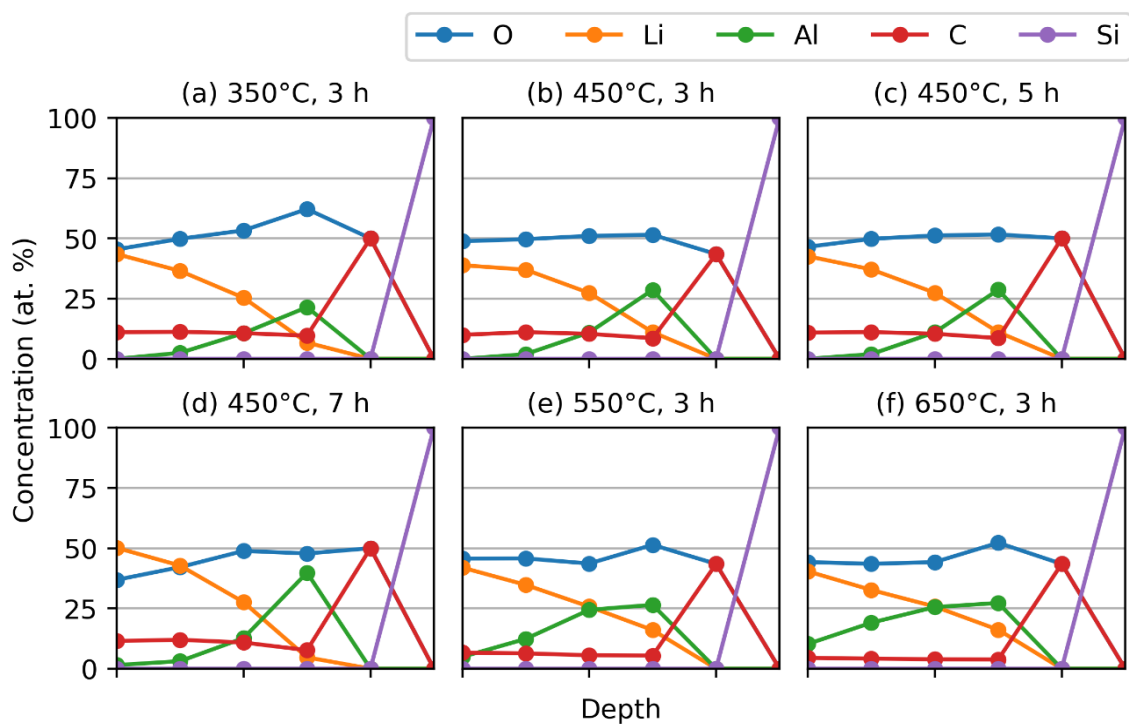

Figure 5: Compositional depth profiles of Li-Al-O thin films after different annealing treatments, determined by D-NRA and RBS.

As shown in Fig. 6 a, the TEM investigations of sample fragment VII in its initial state showed the film to be 320 nm thick. The cross-sectional area from the surface of the film to a depth of approximately 100 nm was very sensitive to the electron beam. Its appearance changed from dark to bright within seconds of exposure to the electron beam (bright areas in Fig. 6a) and could not be examined at high magnification. This depth-dependent sensitivity of the film to the electron beam correlates to the inhomogeneous depth profile observed with XPS (Fig. 4 i), indicating the Li-rich region to be most fragile. The remaining cross-section was found to be amorphous, with very few crystalline grains in the nm size range. An example can be seen in Fig. 6 b. However, low sample conductivity prevented higher quality images. The electron diffraction experiments on the crystalline grains were inconclusive because the diffraction intensity was too low.

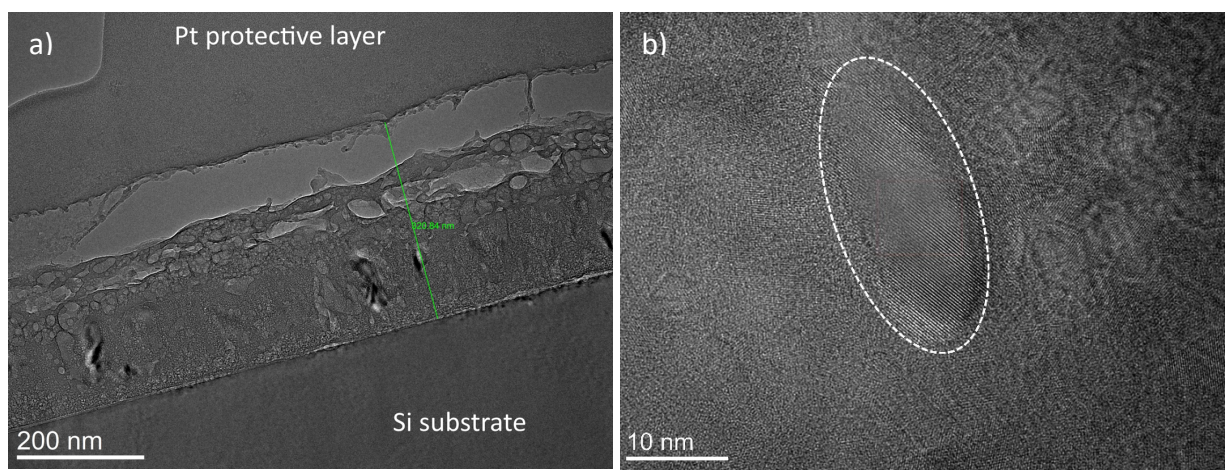

Figure 6: a) Cross-sectional TEM image of a Li-Al-O film. The green line indicates the film thickness of 320 nm. The bright Li-rich areas of the film were affected by the electron beam. b) TEM image of that film shown in a) with higher magnification, including a region with crystalline structure marked with the dashed ellipse.

The observed segregation of Li toward the surface is assumed to be driven by the volatility of Li. There are a few mentions in the literature about the high volatility of Li leading to Li loss by sublimation at elevated temperatures. For example, $LiAlO_2$ powder was observed to transform into $LiAl_5O_8$ above 1200°C, which was attributed to Li sublimation [2]. Yi et al. report severe Li loss from $Li_7La_3Zr_2O_{12}$ nanoparticles, noting that high surface-to-volume ratios (as in nanoparticles and thin films) are particularly challenging [26]. One approach to address Li loss is to use excess Li in the precursor materials [10, 19, 26], although this has the drawbacks of wasting some Li and difficulty in controlling the precise Li content in the final samples. As another approach, Wang et al. developed an ultrafast sintering method and reported Li losses of less than 4 % compared to up to 99 % with conventional sintering methods [27]. Most other publications do not mention Li segregation or sublimation, which is particularly surprising when the reported synthesis routes rely on thermal processing with quite high temperatures and long durations (e.g. up to 877°C for 48 h in [16]). A possible explanation is that in many cases the Li content of bulk/powder/thin film Li-Al-O samples is not actually quantified, but only the crystal structure of Li-Al-O compounds is identified by X-ray diffraction, or the expected phase transitions are observed by differential thermal analysis, as in [11,16]. In other cases (e.g. [22]), surface sensitive methods are used for Li quantification, measuring only the Li-containing near-surface areas, while possibly missing the Li-depleted volumes.

Fig. 7 shows example XRD patterns measured after the different annealing treatments. The high-throughput XRD mapping of the ML in its initial state did not show any diffraction peaks corresponding to crystalline Li-Al-O phases. Similarly, the annealing treatments at 350 – 450°C for 3 – 7 h and 550°C for 7 h did not lead to the formation of crystalline Li-Al-O phases. After the annealing treatments at 550 and 650°C for 3 h, the XRD patterns show low intensity diffraction peaks corresponding to tetragonal γ-LiAlO$_2$ (P4$_1$2$_1$2, ICSD 430357), indicating few crystalline γ-LiAlO$_2$ grains in the otherwise X-ray amorphous films. The RTA annealing treatments at 750 – 850°C for 1 – 10 min resulted in the XRD patterns with the most intense γ-LiAlO$_2$ peaks compared to the other samples. This shows that the minimal temperature for the formation of crystalline LiAlO$_2$ grains was 550°C, but the inhomogeneous depth profiles caused by conventional annealing prevented well-crystallized films. On the other hand, the more homogeneous depth profiles achieved by RTA showed stronger XRD peaks, indicating better crystallinity. While sputtered Li-Al-O films were reported to be amorphous in the as-deposited state [21, 22] and crystalline (γ-LiAlO$_2$) after 2 h annealing at 950°C [21], we identified 550°C as the minimum temperature for γ-LiAlO$_2$ formation.

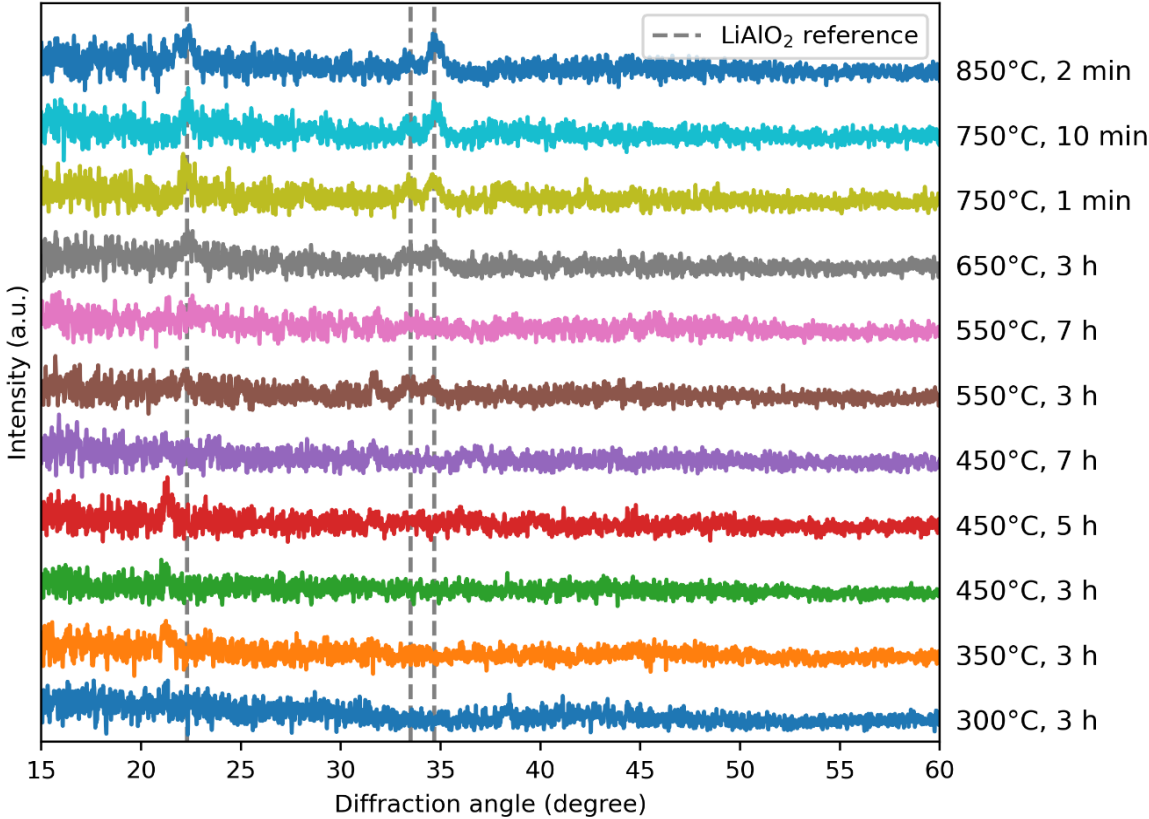

Figure 7: Stacked XRD patterns of Li-Al-O thin films measured after different annealing treatments. The vertical dashed lines indicate the reference diffraction angles of tetragonal γ-LiAlO$_2$ (P4$_1$2$_1$2, ISCD 430357).

**Conclusions**

Annealing temperature and duration have a large influence on the compositional depth profiles and crystallinity of sputtered Li-Al-O thin films. For the formation of LiAlO$_2$, the X-ray amorphous Li-Al-O thin films required a minimum annealing temperature of 550°C. However, conventional annealing (300 – 650°C, 3 – 7 h) also resulted in inhomogeneous depth profiles, with Li contents increased toward the film surface and Al contents toward the film-substrate interface, which prevented well-crystallized Li-Al-O films. Rapid thermal annealing (750 – 850°C, 1 – 10 min) was able to even out the inhomogeneous depth profiles and induce the formation of crystalline LiAlO$_2$. Therefore, rapid thermal annealing is recommended over conventional annealing for the thermal processing of Li-containing thin films. These results also demonstrate the importance of appropriate methods for compositional characterization. For the most accurate characterization without missing segregation effects, depth sensitive methods must be used. Attention must also be paid to the choice of substrate material, as Li was observed to react with the Si substrate at temperatures of 550°C or higher.

**Declaration of competing interest**

The authors declare that they have no known competing financial interests or personal relationships that could have appeared to influence the work reported in this paper.

**Acknowledgements**

This research was funded by DFG SPP 2315, project number 470309740: LU 1175/36-1. The Center for Interface Dominated High-Performance Materials (ZGH, Ruhr University Bochum) is acknowledged for access to XRD, FIB and TEM. The RUBION (Ruhr University Bochum) is acknowledged for D-NRA and RBS measurements.